\documentclass[sigconf]{acmart}
\AtBeginDocument{%
  }

\setcopyright{acmlicensed}
\copyrightyear{2018}
\acmYear{2018}
\acmConference[KDD '26]{Make sure to enter the correct
  conference title from your rights confirmation email}{August 09--13,
  2026}{Jeju, Korea}

\usepackage{tikz}
\usetikzlibrary{positioning, arrows.meta, shapes, calc, fit, backgrounds}
\definecolor{seqFill}{RGB}{154,197,205}
\definecolor{seqLine}{RGB}{34,120,155}
\definecolor{graphFill}{RGB}{240,110,110}
\definecolor{graphLine}{RGB}{195,60,60}
\definecolor{alignFill}{RGB}{190,192,231}
\definecolor{seqTag}{RGB}{214,224,190}
\definecolor{btDark}{RGB}{35,60,170}
\definecolor{btMid}{RGB}{110,140,220}
\definecolor{btLight}{RGB}{205,215,245}
\begin{document}

\title{Cross-Representation Knowledge Transfer for Improved Sequential Recommendations}

\author{Artur Gimranov}
\authornote{Authors contributed equally to this research.}
\email{agimranov@hse.ru}
\orcid{0009-0006-4296-3165}
\affiliation{%
  \institution{HSE University}
  \city{Moscow}
  \country{Russian Federation}
}

\author{Viacheslav Yusupov}
\email{vyusupov@hse.ru}
\orcid{0009-0007-4087-721X}
\authornotemark[1]
\affiliation{%
  \institution{HSE University}
  \city{Moscow}
  \country{Russian Federation}
}

\author{Elfat Sabitov}
\email{a.sabitov@hse.ru}
\orcid{0009-0002-3174-0565}
\authornotemark[1]
\affiliation{%
  \institution{HSE University}
  \city{Moscow}
  \country{Russian Federation}
}

\author{Tatyana Matveeva}
\email{tpmatveeva@hse.ru}
\orcid{0009-0004-0680-5050}
\authornotemark[1]
\affiliation{%
  \institution{HSE University}
  \city{Moscow}
  \country{Russian Federation}
}

\author{Anton Lysenko}
\email{av.lysenko@hse.ru}
\orcid{0000-0002-9369-7104}
\affiliation{%
  \institution{HSE University}
  \city{Moscow}
  \country{Russian Federation}
}

\author{Ruslan Israfilov}
\orcid{0000-0002-4226-9644}
\affiliation{%
  \institution{SB AI Lab}
  \city{Moscow}
  \country{Russian Federation}
}

\author{Evgeny Frolov}
\orcid{0000-0003-3679-5311}
\affiliation{%
  \institution{AXXX}
  \institution{HSE University}
  \city{Moscow}
  \country{Russian Federation}
}

\renewcommand{\shortauthors}{Gimranov et al.}

\begin{abstract}
  Transformer architectures, capable of capturing sequential dependencies in the history of user interactions, have become the dominant approach in sequential recommender systems. Despite their success, such models consider sequence elements in isolation, implicitly accounting for the complex relationships between them. Graph neural networks, in contrast, explicitly model these relationships through higher order interactions but are often unable to adequately capture their evolution over time, limiting their use for predicting the next interaction. To fill this gap, we present a new framework that combines transformers and graph neural networks and aligns different representations for solving next-item prediction task. Our solution simultaneously encodes structural dependencies in the interaction graph and tracks their dynamic change. Experimental results on a number of open datasets demonstrate that the proposed framework consistently outperforms both pure sequential and graph approaches in terms of recommendation quality, as well as recent methods that combine both types of signals.
\end{abstract}



\begin{teaserfigure}
  \centering
  \resizebox{1.00\textwidth}{!}{%

\begin{tikzpicture}[
  /utils/exec={
    \definecolor{seqFill}{RGB}{225, 240, 255}
    \definecolor{seqLine}{RGB}{70, 130, 180}
    \definecolor{seqTag}{RGB}{230, 255, 230}
    \definecolor{seqBlockFill}{RGB}{100, 200, 100}
    \definecolor{graphFill}{RGB}{255, 230, 230}
    \definecolor{graphLine}{RGB}{200, 80, 80}
    \definecolor{alignFill}{RGB}{230, 230, 250}
    \definecolor{layerInner}{RGB}{255, 180, 180}
    \definecolor{layerOuter}{RGB}{255, 220, 220}
  },
  font=\Large\sffamily\bfseries,
  >={Latex[length=3mm]},
  node distance=10mm and 15mm,
  every node/.style={align=center, font=\Large\bfseries},
  box/.style={rectangle, draw, thick, rounded corners=1pt, outer sep=0pt},
  seq_block/.style={box, fill=seqFill, draw=seqLine},
  graph_block/.style={box, fill=graphFill, draw=graphLine},
  input_id/.style={box, fill=white, minimum width=1.2cm, minimum height=2.5cm, font=\Large\bfseries},
  embedding/.style={box, minimum width=1.2cm, minimum height=3.0cm},
  encoder/.style={box, minimum width=5.5cm, minimum height=3.8cm, font=\Large\bfseries},
  loss/.style={box, minimum width=1.2cm, minimum height=3.2cm},
  conn/.style={->, thick, rounded corners=5pt},
  dconn/.style={->, thick, dashed, shorten >=2pt, shorten <=2pt},
  scheme_circle/.style={circle, draw, thick, minimum size=3.8cm, align=center},
  item_node/.style={circle, draw, thick, fill=white, minimum size=0.9cm, font=\small, inner sep=0pt},
  graph_node/.style={circle, draw, thick, fill=white, minimum size=0.70cm, font=\scriptsize\bfseries, inner sep=0pt},
]

\node[box, draw=seqBlockFill, fill=seqTag, minimum width=1.4cm, minimum height=3.2cm] (seq) {\rotatebox{90}{Sequence}};

\node[input_id, draw=seqLine, above right=1.5cm and 2.0cm of seq.east] (itemid) {\rotatebox{90}{Item ID}};
\node[embedding, draw=seqLine, fill=seqFill!30, right=of itemid] (itememb) {\rotatebox{90}{Item Emb.}};
\node[encoder, seq_block, right=of itememb] (seqenc) {Sequential\\Encoder};
\node[loss, draw=seqLine, fill=seqFill!60, right=of seqenc] (locloss) {\rotatebox{90}{Local Loss}};

\node[input_id, draw=graphLine, below right=1.5cm and 2.0cm of seq.east] (userid) {\rotatebox{90}{User ID}};
\node[embedding, draw=graphLine, fill=graphFill!30, right=of userid] (useremb) {\rotatebox{90}{User Emb.}};
\node[encoder, graph_block, right=of useremb] (graphenc) {Graph\\Encoder};
\node[loss, draw=graphLine, fill=graphFill!60, right=of graphenc] (globloss) {\rotatebox{90}{Global Loss}};

\draw[conn] (seq.east) -- ++(1.0,0) |- (itemid.west);
\draw[conn] (seq.east) -- ++(1.0,0) |- (userid.west);
\draw[conn] (itemid) -- (itememb); \draw[conn] (itememb) -- (seqenc); \draw[conn] (seqenc) -- (locloss);
\draw[conn] (userid) -- (useremb); \draw[conn] (useremb) -- (graphenc); \draw[conn] (graphenc) -- (globloss);

\draw[conn] (itememb.south) -- ++(0,-0.8) -| ($(graphenc.west)+(-0.5,0.5)$) -- ($(graphenc.west)+(0,0.5)$);

\coordinate (midEnc) at ($(seqenc.east)!0.5!(graphenc.east)$);

\node[box, fill=alignFill, draw=alignFill!60!blue, minimum width=6.0cm, minimum height=1.5cm, 
      right=2.5cm of midEnc] (alignloss) 
      {\textbf{Representation Alignment Loss}};

\draw[conn] (seqenc.east) -- ++(0.5,0) |- (alignloss.west);
\draw[conn] (graphenc.east) -- ++(0.5,0) |- (alignloss.west);

\node[scheme_circle, draw=seqLine, above=0.3cm of alignloss] (seq_scheme) {};
\node[above, font=\Large\bfseries] at (seq_scheme.north) {Sequential Encoder Scheme};

\begin{scope}[shift={($(seq_scheme.center)+(-0.25cm,0)$)}]
    \node[item_node, draw=seqLine] (u) at (-1.1, 0.5) {User};
    \node[item_node, draw=seqLine] (t1) at (-0.5, -0.6) {Item\\$t-1$};
    \node[item_node, draw=seqLine] (t) at (0.8, 0.5) {Item\\$t$};
    \node[item_node, draw=seqLine] (t2) at (1.6, -0.6) {Item\\$t+1$};
    \draw[->, >={Stealth[length=3mm]}, very thick, draw=seqLine] (u) -- (t1);
    \draw[->, >={Stealth[length=3mm]}, very thick, draw=seqLine, bend left] (t1) to (t);
    \draw[->, >={Stealth[length=3mm]}, very thick, draw=seqLine, bend left] (t) to (t2);
\end{scope}

\node[circle, draw=graphLine, fill=layerOuter, minimum size=3.8cm, below=0.3cm of alignloss] (graph_scheme) {};
\node[circle, draw=graphLine, fill=layerInner, minimum size=1.9cm] (inner_layer) at (graph_scheme.center) {};
\node[below=0.2cm, font=\Large\bfseries] at (graph_scheme.south) (graph_label) {Graph Encoder Scheme};

\begin{scope}[shift={(graph_scheme.center)}]
    \node[graph_node, draw=graphLine, fill=white] (center_u) at (0,0) {User};
    \node[graph_node, draw=graphLine] (in1) at ( 45:0.9cm) {Item};
    \node[graph_node, draw=graphLine] (in2) at (135:0.9cm) {Item}; 
    \node[graph_node, draw=graphLine] (in3) at (225:0.9cm) {Item};
    \node[graph_node, draw=graphLine] (in4) at (315:0.9cm) {Item};
    \node[graph_node, draw=graphLine] (out1) at ( 90:1.6cm) {User};
    \node[graph_node, draw=graphLine] (out2) at ( 30:1.6cm) {User};
    \node[graph_node, draw=graphLine] (out3) at (350:1.6cm) {User};
    \node[graph_node, draw=graphLine] (out5) at (250:1.6cm) {User};
    \node[graph_node, draw=graphLine] (out6) at (180:1.6cm) {User};
    
    \draw[conn, -] (center_u) -- (in1); \draw[conn, -, dashed] (center_u) -- (in2); 
    \draw[conn, -] (center_u) -- (in3); \draw[conn, -] (center_u) -- (in4);
    \draw[conn, -] (in1) -- (out1); \draw[conn, -] (in1) -- (out2);
    \draw[conn, -] (in2) -- (out6); \draw[conn, -] (in2) -- (out1); 
    \draw[conn, -] (in3) -- (out6); \draw[conn, -] (in3) -- (out5);
    \draw[conn, -] (in4) -- (out3);
\end{scope}

\draw[dconn] (seqenc.north) to[bend left=35] (seq_scheme.150);
\draw[dconn] (graphenc.south) to[bend right=35] (graph_scheme.210);

\begin{scope}[on background layer]
    \node[
        draw=black!60, 
        dashed, 
        thick, 
        fill=black!2,
        inner sep=15pt, 
        rounded corners=10pt,
        fit=(userid) (useremb) (graphenc) (globloss) (graph_scheme) (graph_label)
    ] (warmup) {};
\end{scope}

\node[anchor=south west, font=\huge\bfseries\itshape, text=black!70] 
    at (warmup.south west) {Warm-up Stage};

\end{tikzpicture}
  }
\caption{Overview of the CREATE framework. A sequential encoder (top) models users’ local interaction dynamics, while a graph encoder (bottom) captures global item–item context. Their embeddings are aligned via a representation-alignment objective. Training proceeds in two phases: the graph encoder is first pre-trained during a warm-up stage, after which both encoders are jointly optimized end-to-end.}
  \Description{Block diagram of CREATE with two encoders (sequential and graph) whose embeddings are aligned; the graph branch is pre-trained in a warm-up stage before joint training.}
  \label{fig:create}
\end{teaserfigure}
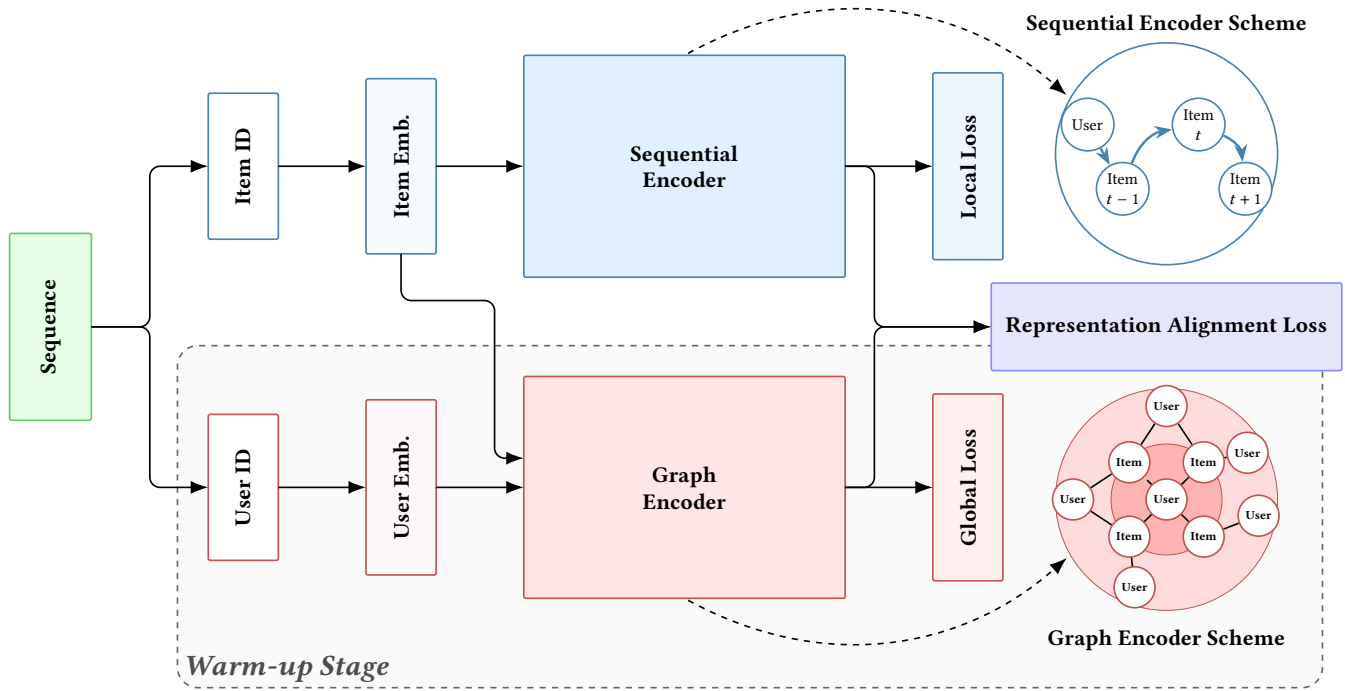


\maketitle

\section{Introduction}
Recommender systems play a crucial role in modern digital platforms. They leverage available data on user--item interactions to rank items that a user is likely to prefer in the future. Recently, deep learning-based approaches have become increasingly popular~\cite{zhang2019deep,gheewala2025depth} dominating recommendation systems, particularly in the area of sequence-based recommendations~\cite{yoon2023evolution} but not limited to them \cite{he2020lightgcn,zhang2024transgnn,chen2025lightgnn}.

For recommendations on sequences where user interactions are modeled over time, state-of-the-art models, such as SASRec \cite{sasrec} and BERT4Rec \cite{bert4rec}, primarily utilize users' collaborative signal, considering each sequence of user items independently. This allows the model to capture the relationships between items based on their order, and build rich representations of both users and items. However, such models do not account for certain global relationships of items and users between each other.

However, depending on the domain, not all data is sequential \cite{klenitskiy2024does}, and such scenarios may require models that do not rely on the order of interactions. One such approach is the use of graph neural networks (GNNs), which model user-item interaction graph. Recent studies demonstrate the application of graph neural networks in recommendation tasks~\cite{wu2022graph,anand2025survey}, showing superiority over non-sequential models. However, GNNs do not account for the sequential nature of interactions, resulting in inferior recommendation quality compared to SOTA sequence-based models~\cite{luo2024collaborative}. Thus, both sequence and graph approaches attempt to model user preferences by considering different data structures: local (a user's item sequence), and global context, expressed as a user subgraph, explicitly defining relationships between users and items.

Recent work has shown that using different interaction structures can improve the quality of recommendations in the next-item prediction task~\cite{mrgsrec,gsau,loom}. These studies combine two architectures: transformers and GNNs, where SASRec is used most often as the transformer model, and LightGCN \cite{he2020lightgcn} or other graph encoders are used as the graph model. Despite this, existing solutions have a number of significant drawbacks related to the way the resulting representations are fused, as well as the method for predicting new recommendations.

In our work, we present a Cross-REpresentation Aligned Transfer Encoders (CREATE) framework that combines transformer and graph architectures in a way that enriches the sequence-based model with global knowledge and produces representations aligned with each other (see Figure \ref{fig:create}). It mitigates the problems of previous works: (i) recommendation updates are based on the sequential encoder's scheme, which avoids usage of user embeddings, making our approach asymmetric and independent from folding-in procedure \cite{yusupov2025ultra}, an approach for updating user embeddings in correspondence to new interactions; 
(ii) for representation alignment, Barlow Twins is used instead of contrastive loss, which shows better results and enables redundancy reduction. Thus, our contributions are:

\begin{itemize}
    \item We propose a framework that combines transformer and graph architectures for the next item prediction task;
    \item We develop a new method of training a model under the framework that helps improve the quality of recommendations;
    \item We introduce a representation alignment method between encoders, which enables redundancy reduction and improves recommendation quality.
\end{itemize}

All the experiments are implemented in a global temporal split setting, which is a better approximation to the industrial setting than leave-last-out \cite{gusak2025time}. The rest of the paper is organized as follows: Sec. \ref{related} contains detailed related work, Sec. \ref{approach} proposes our approach, Sec. \ref{exp} describes the experimental setup and results, and Sec. \ref{ablation} presents the ablation study. Code implementation is publicly available at \url{https://anonymous.4open.science/r/multirepr_recsys-761F}.

\section{Related work}
\subsection{Sequential recommendations}
\label{related}
Transformer-based deep models for sequential recommendation \cite{boka2024survey} have become the dominant approach in the field due to their natural suitability for sequential data \cite{klenitskiy2024does}. One of such model is SASRec \cite{sasrec}, offers state-of-the-art performance. However, its original version uses binary cross-entropy as a loss function, which can lead to biased predictions and subsequent studies show that switching to full cross-entropy significantly improves results \cite{scalablece,klenitskiy2023turning,gusak2024rece}. Because SASRec operates solely on the relative order of items in a user's sequence, modifications have been proposed that enrich the model with additional information, such as time intervals between interactions \cite{li2020time}, content \cite{lichtenberg2025denserec,ho2024self} and context \cite{yuan2025contextual,rashed2022context}. Another modern standard is a BERT4Rec \cite{bert4rec}, which in contrast to unidirectional attention, uses a bidirectional attention mechanism, masking out positive items from a sequence. Further developments to transformer-based architectures are aimed at applying self-supervised learning methods to the obtained representations: CL4SRec \cite{xie2022contrastive} uses contrastive learning for this purpose, while the Barlow Twins approach \cite{razvorotnev2025barlow} achieves similar goals without the need for negative sampling.

\subsection{Graph-based recommendations}
Graph neural networks (GNNs) have become a powerful tool in this field because they directly exploit graph structure of user-item interactions \cite{wu2022graph}. This allows them to capture complex, high-level relationships and transitive dependencies that are difficult to detect when examining interactions sequentially or in isolation \cite{anand2025survey,zare2025dac,yu2023xsimgcl,gao2023survey}. Early studies applying GNNs to recommendation problems considered applying existing architectures to a bipartite graph of user-item interactions \cite{wu2022graph}, including NGCF \cite{wang2019neural} and an attention version of graph neural network GAT \cite{kim2025graph}. Later work demonstrated that the general GCN \cite{kipf2016semi} architecture itself can be excessive for recommendation systems, as reflected in LightGCN \cite{he2020lightgcn}, which eschews fully connected neural networks for processing neighbor representations and instead trains only user and item embeddings, significantly improving recommendation quality. Later work UltraGCN \cite{mao2021ultragcn} addresses the oversmoothing problem of a LightGCN by considering user and item representations in the limit, with an infinite number of hops on a graph. Some studies also consider item-item graphs \cite{mao2021ultragcn,loom}, the edges of which are constructed from data or learned during model training.

\subsection{Multi-representational models}
Multi-representational models \cite{du2024disentangled,zhang2024temporal} are defined as models that use object representations from different sources (in our case, embeddings obtained using a sequential encoder and a graph encoder). Transformer architectures, as well as graph architectures, have fundamental limitations. Transformer architectures consider item sequences independently, explicitly failing to consider higher-order relationships. While addressing this shortcoming \cite{yang2023dgrec}, GNNs fail to account for the sequential nature of the data, which clearly impacts the performance of next-item prediction tasks \cite{loom, wang2022multi}. Given this, combining these two types of information is a relevant approach that potentially improves recommendation quality, as confirmed by several studies described below.

Currently, there are few studies combining transformer and graph representations, as the field is new. Among the first studies, the MCLSR \cite{wang2022multi,loom} approach is noteworthy, implementing late fusion of representations obtained from sequential and graph parts, using a contrastive loss to align the representations. Another work that stands out among the late fusion approaches is MRGSRec \cite{mrgsrec}, which, unlike the previous work, uses graph representations during prediction. 
In contrast, some works use an early fusion mechanism. For instance, GSAU \cite{gsau} employs a shared embedding layer for both the transformer and graph components. It uses a contrastive loss function, designed to bring similar items closer and push dissimilar ones apart in the latent space \cite{jing2023contrastive}, and the binary cross-entropy. However, later works demonstrates that replacing BCE loss with cross-entropy in transformer models improves performance \cite{scalablece,klenitskiy2023turning}.
The most recent model of this type is LOOM \cite{loom}, in which a sequential encoder serves as a teacher for a graph using item-item relationships, the weights of which are learned from the data. LOOM also utilizes late fusion of the embeddings obtained by the sequential encoder and the graph model. Here, information enrichment occurs unilaterally, and the user's graph representation is not updated when new data arrives, which may impact the quality of recommendations.

It is also worth noting that all the aforementioned works evaluated their quality using a next-item prediction approach using a leave-last-out partitioning scheme, which differs from the real-world scenario \cite{gusak2025time}.

\section{Proposed approach}
This section describes the problem statement of a sequential recommendations and the proposed cross-representation knowledge transfer framework, which scheme is presented at Figure \ref{fig:create}.

\subsection{Problem statement}
\label{approach}
Let us first denote the set of users as $\mathcal{U}$, and the set of items as $\mathcal{I}$, where $|\mathcal{U}| = M$ and $|\mathcal{I}| = N$. The next-item prediction problem is formulated as predicting the next item $i_{|H_u|+1}^u$ for user $u \in \mathcal{U}$, given his history of past interactions $H_u = (i_1^u, i_2^u, \dots, i_{|H_u|}^u)$.

However, existing recommendation algorithms that use the graph structure of user interactions do not take into account the temporal order of items in the user's history, but rely on so-called global connections between users and items. By this, we can reformulate the problem for graph recommenders as: for a user $u \in \mathcal {U}$, predict the next item $i_{|H_u|+1}^u$ based on user context $\mathcal {C}^u$, including the edges of a bipartite interaction graph $\mathcal {G}$ between users $\mathcal {U}$ and items $\mathcal {I}$.

Below are the components of our CREATE framework, which includes a shared embedding layer, sequential and graph encoders and the alignment component. The overall algorithm diagram is shown in Figure \ref{fig:create}.

\subsection{Embedding layer}
\label{model:embeddings}

We map user and item identifiers into a \(d\)-dimensional space using learnable lookup functions
\[
U_e:\mathcal{U}\to\mathbb{R}^d,\qquad I_e:\mathcal{I}\to\mathbb{R}^d.
\]
Thus, the representations of a user $u$ and an item $i$ are:
\[
\mathbf{e}^u = U_e(u)\in\mathbb{R}^d, \qquad \mathbf{e}_i = I_e(i)\in\mathbb{R}^d,
\]
the user's interaction history \(H_u=(i_1^u,\ldots,i_c^u)\) can be mapped into latent space:
\[
\mathbf{E}_u = (\mathbf{e}_{i_1^u},\ldots,\mathbf{e}_{i_c^u})\in\mathbb{R}^{c\times d},\qquad \mathbf{e}_{i_k^u} = I_e(i_k^u),
\]

Inheriting the logic of transformers, positional embeddings are used to encode ordinal information:
\[
P_e:\{1,\ldots,c\}\to\mathbb{R}^d,
\]
and form position-aware item representations by additive composition:
\[
\hat{\mathbf{e}}_{i_k^u} = \mathbf{e}_{i_k^u} + P_e(k),\qquad
\hat{\mathbf{E}}_u=(\hat{\mathbf{e}}_{i_1^u},\ldots,\hat{\mathbf{e}}_{i_c^u})\in\mathbb{R}^{c\times d}.
\]
This choice is standard in sequential recommenders with learnable positional encodings~\cite{sasrec,bert4rec}.


\subsection{Sequential encoder}

As the sequential encoder in \textsc{CREATE}, we can plug in a broad class of sequence models, including
RNN/CNN-based architectures and, more commonly, Transformer-based recommenders. In this work, we utilized a widely used SASRec encoder with cross-entropy loss \cite{scalablece,klenitskiy2023turning}, which shows the state-of-the-art performance on a broad range of sequential datasets \cite{klenitskiy2024does}. In addition, we also conducted experiments with \textsc{BERT4Rec}~\cite{bert4rec} as a sequential backbone. Where the full CE is not applicable due to the large catalog size, we utilize scalable cross-entropy loss \cite{scalablece} to reduce the computational cost of the model. 

\paragraph{SASRec.}
For each user \(u\), we take the learnable item embeddings from the shared embedding layer (augmented with positional information) and feed them into a causal Transformer encoder, in our case this is SASRec. The uni-directional self-attention
ensures that the representation at position \(t\) only attends to items \(1,\ldots,t\), matching the next-item prediction setup. Denote the encoder output by
\[
\mathbf{H}_u^\textsc{sasrec} = \mathrm{TransformerEncoder}\!\left(\hat{\mathbf{E}}_u\right) \in \mathbb{R}^{c \times d},
\]
and let \(\mathbf{h}_u \in \mathbb{R}^{d}\) be the user-state vector used for next-item prediction (e.g., the output at the last position):
\[
\mathbf{h}_u = \mathbf{H}_u^{\textsc{sasrec}}[c].
\]


\paragraph{BERT4Rec.}
 Alternatively, \textsc{BERT4Rec}~\cite{bert4rec} uses a bidirectional Transformer encoder trained via masked item prediction. Unlike causal masking, the self-attention mechanism has access to all positions in the sequence, using both past and future interactions. During training, we randomly select a subset of positions \( \mathcal{M} \subset \{1,\ldots,c\} \) and replace their corresponding item embeddings with a learned \texttt{[MASK]} embedding \(\mathbf{e}_{\textsc{mask}} \in \mathbb{R}^d\). The bidirectional encoder processes the entire modified sequence:
\[
\mathbf{H}_u^{\textsc{bert}} = \mathrm{BidirectionalTransformer}\!\left(\tilde{\mathbf{E}}_u\right) \in \mathbb{R}^{c \times d},
\]

At inference time, we append an extra \texttt{[MASK]} token at last position and take its output representation as the user-state vector:
\[
\mathbf{h}_u = \mathbf{H}_u^{\textsc{bert}}[c].
\]

\paragraph{Prediction head and training objective.}
Given a predicted user-state vector \(\mathbf{h}_u\) from a transformer, we compute
scores for all items via a dot product with the item embedding matrix \(\mathbf{W}_I \in \mathbb{R}^{N \times d}\)
(typically the same learnable embedding table as \(I_e\)):
\[
\mathbf{s}_u = \mathbf{W}_I \mathbf{h}_u \in \mathbb{R}^{N},
\qquad
s_{u,i} = \langle \mathbf{h}_u, \mathbf{e}_i \rangle,
\]
where \(\mathbf{e}_i\) is the embedding of item \(i\).

For training, we use the full softmax cross-entropy objective over the entire item set:
\[
\mathcal{L}_{\mathrm{full}} =
-\sum_{(u,t)} \log
\frac{\exp\left(s_{u,i^{u}_{t+1}}\right)}
{\sum_{j \in \mathcal{I}} \exp\left(s_{u,j}\right)},
\]
where \((u,t)\) ranges over training positions, and \(i^{u}_{t+1}\) denotes the ground-truth next item.


\subsection{Graph encoder}
The graph backbone in CREATE operates on the user-item graph of interactions, sharing item embeddings with the sequential backbone, capturing global interaction patterns. In particular, user embeddings are utilized only during training; inference relies solely on the sequential backbone with enriched item embeddings. This design eliminates the need for a "folding-in" procedure \cite{yusupov2025ultra} to update user representations for new interactions. Consequently, CREATE remains robust to out-of-training user activity and supports asymmetric graph structures \cite{asymmetrichyperbolic} that prioritize item-item relations over persistent user states.

As graph backbone we used LightGCN \cite{he2020lightgcn} and UltraGCN \cite{mao2021ultragcn}. LightGCN is a simplified Graph Convolution model, which omits operations such as self-connection, feature transformation, and nonlinear activations. UltraGCN further simplifies Graph Convolution and considers the limit case of message passing for an infinite amount of layers.

Regarding the training process, LightGCN optimizes the Bayesian Personalized Ranking (BPR) loss \cite{BPRMF}: the difference between positive $s^u_{i \in H_u}$ and negative scores $s^u_{j \notin H_u}$ of the graph model
\[L_{BPR} = -\sum_{u\in \mathcal U} \sum_{i \in H_u} \sum_{j \notin H_u} \ln \sigma (s^u_{i} - s^u_{j}).\]
Optionally, the regularization term can be added $\lambda_{reg} (\sum_{u\in \mathcal U}\|\mathbf e^u\| + \sum_{u\in \mathcal U,\, i \in H_u} \|\mathbf e_i\|)$, which regulates the $L2$ norms of the user and item embeddings, respectively.

The UltraGCN model \cite{mao2021ultragcn} optimizes the loss function consisting of three parts:

\begin{equation}
    L = L_O + \lambda L_C + \gamma L_I,
\end{equation}

where $L_O$ stands for the general objective, which is Binary Cross-Entropy (BCE) for link prediction with uniform weighting and hard negatives; $L_C$ is a constraint component, which helps UltraGCN approximate infinite-layer propagation and $L_I$ stands for item-item constraint.







\subsection{Representation alignment}
\label{subsec:repr-alignment}

Our model forms two views of each user: a sequential (\emph{local}) representation capturing short-range intent, and a graph-based (\emph{global}) representation capturing dependencies induced by the interaction graph. Since both feed a shared scoring function, they should encode a consistent user state.

We enforce consistency with a redundancy-reduction objective that promotes (i) invariance between the two views of the same user and (ii) low feature redundancy in the aligned space. The latter is important in our setting: without decorrelation, the encoders can agree through a few correlated coordinates, reducing effective capacity and destabilizing optimization. We therefore use the Barlow Twins objective~\cite{zbontar2021barlowtwins}, which drives the cross-correlation between paired views toward the identity matrix.

Let \(h_u\) denote the local representation from the sequential encoder and \(e^u\) the global representation from the graph encoder. For a batch \(\mathcal{B}\) of \(B\) users, we standardize both embeddings along the batch dimension (e.g., batch normalization without affine parameters) to obtain \(\mathbf{z}^{\mathrm{loc}}_u\) and \(\mathbf{z}^{\mathrm{glob}}_u\). We then compute the empirical cross-correlation
\begin{equation}
\mathbf{C}_{ij}
=
\frac{1}{B}\sum_{u \in \mathcal{B}} z^{\mathrm{loc}}_{u,i}\, z^{\mathrm{glob}}_{u,j},
\qquad i,j \in \{1,\dots,d\},
\end{equation}
where \(z^{\mathrm{loc}}_{u,i}\) and \(z^{\mathrm{glob}}_{u,j}\) are the \(i\)-th and \(j\)-th components of the standardized embeddings. The Barlow Twins loss is
\begin{equation}
\mathcal{L}_{\mathrm{BT}}
=
\sum_{i=1}^{d}\bigl(1-\mathbf{C}_{ii}\bigr)^{2}
+
\lambda \sum_{\substack{i,j=1\\ i\neq j}}^{d}\mathbf{C}_{ij}^{2},
\end{equation}
where the first term encourages invariance by pushing \(\mathrm{diag}(\mathbf{C})\) toward \(1\), and the second reduces redundancy by pushing off-diagonal correlations toward \(0\); \(\lambda>0\) controls the trade-off. We add this regularizer to the recommendation objective:
\begin{equation}
\mathcal{L}
=
\mathcal{L}_{\mathrm{rec}} + w_{BT}\,\mathcal{L}_{\mathrm{BT}},
\end{equation}
where \(\mathcal{L}_{\mathrm{rec}}\) is the task loss (e.g., next-item prediction / ranking) and \(w_{BT}\) controls alignment strength.

\subsection{Training process}
The essential part in multi-representational setup is to make sure that components do not conflict with each other. Our framework optimizes the weighted loss:

\begin{equation}\label{eq:framework_loss}
    L = L_{local} + w_{global} L_{global} + w_{BT} L_{BT}.
\end{equation}

The first component, $L_{local}$ is the loss function for the sequential model (full cross-entropy in our case) and it serves as the basis. The second term $L_{global}$ is responsible for graph learning and the last one $L_{BT}$ is Barlow Twins for representation alignment.

The last two terms have weights $w_{global}$ and $w_{BT}$ (Eq. \ref{eq:framework_loss}), respectively, so the influence of the graph model during the training process can be adjusted for each dataset individually.

\begin{figure}[H]
    \centering
    \includegraphics[width=0.48\textwidth]{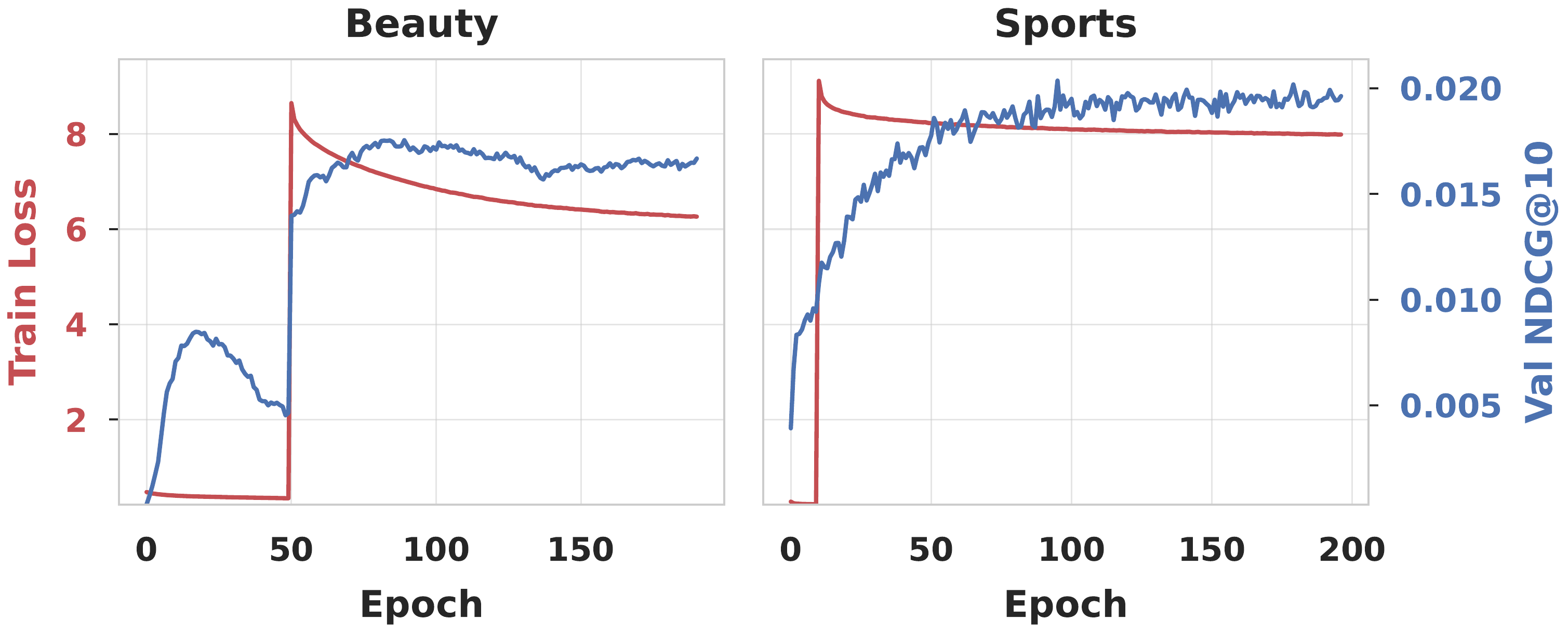}
    \caption{Visualized training process, with training loss and NDCG@10 scales placed on the left and on the right accordingly.}
\Description{Training curves for Beauty and Sports datasets showing distinct warm-up and joint training phases. In both datasets, training loss (red line, left axis) drops sharply after epoch when the sequential model is activated, then gradually decreases. The validation NDCG@10 metric (blue line, right axis) shows consistent improvement, with Beauty reaching ~0.018 and Sports reaching ~0.020, both stabilizing after approximately 50-100 epochs. The sharp transition at the warm-up boundary demonstrates the model successfully integrating graph-based knowledge with sequential patterns.}
    \label{fig:loss_dynamics}
\end{figure}

Our framework relies on the sequential model, enriched by graph representation. To make sure that the graph model provides relevant knowledge, we first let it "warm-up". To do that, during the first $N_{warmup}$ epochs, we train only the graph model. After that, it produces better item embeddings, so we turn on the sequential model and train the whole model. This procedure increases overall performance, which can be seen clearly in the corresponding ablation section. The dynamics of the training process is shown in the Figure~\ref{fig:loss_dynamics}.

\section{Experiments}
In this section, we present the main experimental results of our work and describe the evaluation protocol, including the datasets and training setup. We compare our method against standard Transformer-based baselines, recent multi-representational models, and classical non-neural recommenders. The complete results are reported in Table~\ref{table:comparison_full}.
\subsection{Datasets}
\label{exp}
We benchmark recommendation models on MovieLens-1M\footnote{\url{https://grouplens.org/datasets/movielens/1m}}, three Amazon product categories, and the Yambda music dataset. The statistics of datasets could be seen in the Table \ref{tab:stats}. For MovieLens and Amazon, we treat an \emph{interaction} as an explicit $1$--$5$ rating associated with a timestamp. 
For Yambda, we treat an interaction as a timestamped user action on a track (including both implicit and explicit feedback), following the dataset specification.

For e-commerce recommendation, we use the Amazon review datasets~\cite{amazon_review_data_jmcauley} (1996--2018) in their 5-core filtered versions, retaining only users and items with at least five interactions; we consider Clothing, Shoes and Jewelry, Sports and Outdoors, and Beauty. Besides ratings and timestamps, Amazon data may include review text and rich item metadata (text, categorical descriptors such as brand and category hierarchy, numerical signals such as price and popularity proxies, and optional related-item links).

Yambda-5B~\cite{yambda} (Yandex Music Billion-interactions DAtaset) is a large-scale open dataset of music consumption collected from Yandex.Music (see Table \ref{tab:stats}). 
The log of the dataset includes five event types: \emph{Listen} (implicit), \emph{Like}/\emph{Dislike} (explicit), and \emph{Unlike}/\emph{Undislike} (preference cancellations), with temporal order preserved at up to 5-second granularity. Yambda further provides the \texttt{is\_organic} flag (organic vs.\ recommender-driven actions), pretrained audio embeddings for tracks (CNN-based, contrastive training on spectrograms), and track/album/artist metadata. The dataset is released in multiple scale variants (e.g., Yambda-50M, Yambda-500M, Yambda-5B) to support experiments under different computational budgets.


\begin{table}[ht]
  \centering
  \caption{Datasets and their statistics used in this work.}
  \label{table:datasets}
  \begin{tabular}{lrrr}
    \toprule
    Dataset & \#Users & \#Items & \#Interactions \\
    \midrule
    MovieLens-1M & 6{,}040 & 3{,}706 & 1{,}000{,}209 \\
    Amazon Clothing (5-core) & 39{,}387 & 23{,}033 & 278{,}677 \\
    Amazon Sports (5-core) & 35{,}598 & 18{,}357 & 296{,}337 \\
    Amazon Beauty (5-core) & 22{,}363 & 12{,}101 & 198{,}502 \\
    Yambda-50M (likes) & 8283 & 181,304 & 881,456 \\
    \bottomrule
  \end{tabular}
  \label{tab:stats}
\end{table}

\subsection{Evaluation protocol}


In this work, we utilize the global time split strategy proposed by \cite{gusak2025time} to partition each recommendation dataset into training, validation, and test sets. All models, including ours and the baselines, are firstly trained on the training set, and their hyperparameters are tuned using the validation set. Finally, each model is retrained on the combined training and validation datasets using its optimal hyperparameters, and all reported metrics are evaluated on the test dataset.

\subsection{Baselines}

We compare our CREATE model with other approaches according to  $NDCG@K$, $Recall@K$ and $Coverage@K$ metrics for $K\in \{10, 100\}$ across all datasets. Our model is compared with traditional baselines, such as SVD and Random; two sequential baselines: SASRec \cite{sasrec} and BERT4Rec \cite{bert4rec}; three graph models: LightGCN \cite{he2020lightgcn}, UltraGCN \cite{mao2021ultragcn} and Self-GNN \cite{liu2024selfgnn}; the latest multi-representational recommender systems, such as LOOM \cite{loom}, MRGSRec \cite{mrgsrec} and GSAU \cite{gsau}. To ensure a consistent comparison with recent multi-representational methods like LOOM and GSAU, we evaluate SASRec and our CREATE approach in both: our setup and the LOOM setup. In these experiments, SASRec is used as a common baseline to bridge the two setups.




\subsection{Performance comparison with state-of-the-art methods}


Table ~\ref{table:comparison_full} demonstrates the comparison between our CREATE approach with diverse set of baselines across five datasets. The gains demonstrate the improvement over strong sequential baseline SASRec \cite{sasrec} with scalable cross-entropy loss \cite{scalablece}. Overall, \textsc{CREATE} consistently improves recommendation quality across most datasets, and the gains remain stable for both cutoffs $K=10$ and $K=100$, with significant improvement in \textbf{+38\%} on NDCG@10 and \textbf{+26\%} on NDCG@100 on Yambda-50M dataset. We find these results especially encouraging, as they suggest that \textsc{CREATE} can deliver strong improvements on data that better reflects practical deployment scenario.

Across all five datasets, \textsc{CREATE} achieves the strongest ranking quality and recall, indicating consistent advantages over sequential-only (BERT4Rec \cite{bert4rec}, SASRec \cite{sasrec}), graph-only (LightGCN \cite{he2020lightgcn}, UltraGCN \cite{mao2021ultragcn}) and multi-representational (MRGSRec \cite{mrgsrec}) architectures (see Table ~\ref{table:comparison_full}). 

Importantly, \textsc{CREATE} improves upon \emph{each} encoder family represented in the table. Compared to sequential baselines, the gains suggest that augmenting sequence modeling with graph connectivity provides complementary signals beyond next-item dynamics alone. Compared to graph-based baselines, the improvements indicate that incorporating temporal user dynamics strengthens orderless graph representations. As expected, Random and PopRnd obtain very high coverage by design, yet they substantially underperform in ranking quality and recall, emphasizing that high exposure alone does not yield accurate personalization.

Overall, these results support the main premise of \textsc{CREATE}: combining sequential and graph encoders leads to a more expressive representation that can improve the performance of both architectural families, often yielding a favorable accuracy--coverage trade-off.




\newcommand{\posg}[1]{\textbf{#1}}
\newcommand{\negg}[1]{{#1}}
\newcommand{\rotationvar}{55}

\begin{table*}[t]
  \caption{Comparison of models, evaluated using $NDCG@K$, $Recall@K$ and $Coverage@K$ (in $\%$), with SASRec underlined as main baseline.}
  \label{table:comparison_full}
  \begin{tabular}{lccccccccccc}
    \toprule
    \rotatebox{\rotationvar}{Dataset} & \rotatebox{\rotationvar}{Metric} & \rotatebox{\rotationvar}{\underline{SASRec}} & \rotatebox{\rotationvar}{BERT4Rec} & \rotatebox{\rotationvar}{MRGSRec} & \rotatebox{\rotationvar}{SVD} & \rotatebox{\rotationvar}{Random} & \rotatebox{\rotationvar}{PopRnd} & \rotatebox{\rotationvar}{LightGCN} & \rotatebox{\rotationvar}{UltraGCN} & \rotatebox{\rotationvar}{\textbf{CREATE}} & \rotatebox{\rotationvar}{Gain}\\
    \midrule
    Beauty & NDCG@10 & \underline{1.87 ($\pm 0.01$)} & 1.34 & 0.41 & 0.52 & 0.04 & 0.07 & 0.93 & 0.97 & 2.15 ($\pm 0.01$) & \posg{+15\%}\\
           & Recall@10 & \underline{3.83 ($\pm 0.02$)} & 2.68 & 0.88 & 1.04 & 0.07 & 0.15 & 1.88 & 1.93 & 4.65 ($\pm 0.03$) & \posg{+21\%}\\
           & Cov@10 & \underline{70.04 ($\pm 1.26$)} & 21.42 & 3.24 & 14.02 & 99.90 & 91.24 & 13.27 & 17.54 & 75.67 ($\pm 1.26$) & \posg{+8\%}\\
           & NDCG@100 & \underline{3.73 ($\pm 0.02$)} & 3.17 & 1.27 & 1.92 & 0.18 & 0.26 & 2.13 & 2.25 & 4.25 ($\pm 0.03$) & \posg{+14\%}\\
           & Recall@100 & \underline{13.67 ($\pm 0.08$)} & 12.19 & 5.38 & 7.22 & 0.82 & 1.21 & 8.25 & 8.34 & 15.64 ($\pm 0.13$) & \posg{+14\%}\\
           & Cov@100 & \underline{93.46 ($\pm 3.29$)} & 54.88 & 21.80 & 72.79 & 100.00 & 96.85 & 46.94 & 47.33 & 95.37 ($\pm 1.04$) & \posg{+2\%}\\
    \midrule
    Clothing & NDCG@10 & \underline{0.87 ($\pm 0.01$)} & 0.99 & 0.51 & 0.57 & 0.01 & 0.03 & 0.78 & 0.79 & 1.00 ($\pm 0.01$) & \posg{+15\%}\\
             & Recall@10 & \underline{1.70 ($\pm 0.02$)} & 1.89 & 1.01 & 1.43 & 0.04 & 0.06 & 1.62 & 1.65 & 2.19 ($\pm 0.02$) & \posg{+29\%}\\
             & Cov@10 & \underline{52.11 ($\pm 3.52$)} & 7.91 & 5.27 & 14.09 & 99.92 & 94.71 & 17.50 & 18.23 & 57.06 ($\pm 3.52$) & \posg{+9\%}\\
             & NDCG@100 & \underline{1.86 ($\pm 0.02$)} & 2.10 & 1.36 & 1.63 & 0.09 & 0.13 & 1.86 & 1.87 & 2.19 ($\pm 0.01$) & \posg{+18\%}\\
             & Recall@100 & \underline{6.88 ($\pm 0.03$)} & 8.12 & 5.50 & 6.76 & 0.42 & 0.65 & 7.27 & 7.29 & 8.37 ($\pm 0.04$) & \posg{+22\%}\\
             & Cov@100 & \underline{84.55 ($\pm 1.75$)} & 31.53 & 36.17 & 60.90 & 100.00 & 98.15 & 51.10 & 52.86 & 88.67 ($\pm 2.05$) & \posg{+5\%}\\
    \midrule
    Sports & NDCG@10 & \underline{1.48 ($\pm 0.01$)} & 1.35 & 0.56 & 0.95 & 0.04 & 0.07 & 1.13 & 1.19 & 1.58 ($\pm 0.01$) & \posg{+7\%}\\
           & Recall@10 & \underline{2.75 ($\pm 0.02$)} & 2.55 & 1.28 & 1.84 & 0.07 & 0.12 & 2.14 & 2.21 & 2.98 ($\pm 0.02$) & \posg{+8\%}\\
           & Cov@10 & \underline{50.35 ($\pm 1.21$)} & 18.36 & 1.57 & 18.34 & 99.96 & 93.43 & 17.49 & 18.23 & 75.90 ($\pm 2.61$) & \posg{+50\%}\\
           & NDCG@100 & \underline{3.04 ($\pm 0.01$)} & 3.01 & 1.35 & 2.32 & 0.11 & 0.27 & 2.55 & 2.61 & 3.24 ($\pm 0.01$) & \posg{+7\%}\\
           & Recall@100 & \underline{10.94 ($\pm 0.04$)} & 11.15 & 5.26 & 7.53 & 0.47 & 1.21 & 8.43 & 8.49 & 11.71 ($\pm 0.06$) & \posg{+7\%}\\
           & Cov@100 & \underline{85.26 ($\pm 0.92$)} & 49.49 & 11.01 & 51.32 & 100.00 & 98.26 & 49.60 & 52.72 & 96.33 ($\pm 0.98$) & \posg{+13\%}\\
    \midrule
    ML-1M & NDCG@10 & \underline{7.96 ($\pm 0.66$)} & 7.35 & 1.43 & 2.78 & 0.12 & 0.46 & 3.21 & 3.45 & 8.99 ($\pm 0.26$) & \posg{+13\%}\\
          & Recall@10 & \underline{13.99 ($\pm 1.08$)} & 13.29 & 3.41 & 6.21 & 0.26 & 0.87 & 6.57 & 6.73 & 15.73 ($\pm 0.39$) & \posg{+12\%}\\
          & Cov@10 & \underline{45.04 ($\pm 1.33$)} & 35.00 & 6.61 & 32.10 & 95.41 & 60.75 & 34.13 & 35.49 & 40.22 ($\pm 0.88$) & \negg{-10\%}\\
          & NDCG@100 & \underline{13.09 ($\pm 0.36$)} & 12.35 & 5.14 & 5.46 & 0.70 & 1.81 & 5.98 & 6.17 & 14.06 ($\pm 0.28$) & \posg{+7\%}\\
          & Recall@100 & \underline{39.93 ($\pm 1.01$)} & 38.67 & 23.01 & 18.67 & 3.50 & 8.31 & 20.37 & 21.19 & 41.55 ($\pm 0.94$) & \posg{+4\%}\\
          & Cov@100 & \underline{83.41 ($\pm 0.69$)} & 72.31 & 26.5 & 73.02 & 100.00 & 82.28 & 57.38 & 61.40 & 78.67 ($\pm 0.81$) & \negg{-5\%}\\
    \midrule
    Yambda-50M & NDCG@10 & \underline{1.11 ($\pm 0.01$)} & 0.39 & 0.20 & 0.42 & 0.0 & 0.04 & 1.18 & 1.19 & 1.54 ($\pm 0.01$) & \posg{+38\%}\\
               & Recall@10 & \underline{1.35 ($\pm 0.02$)} & 0.87 & 0.37 & 0.84 & 0.0 & 0.09 & 1.28 & 1.30 & 1.86 ($\pm 0.02$) & \posg{+37\%}\\
               & Cov@10 & \underline{7.80 ($\pm 0.52$)} & 0.006 & 0.01 & 5.21 & 27.61 & 11.94 & 10.88 & 10.99 & 13.70 ($\pm 0.84$) & \posg{+75\%}\\
               & NDCG@100 & \underline{1.46 ($\pm 0.02$)} & 1.04 & 0.69 & 0.71 & 0.003 & 0.18 & 1.21 & 1.25 & 1.84 ($\pm 0.02$) & \posg{+26\%}\\
               & Recall@100 & \underline{3.15 ($\pm 0.03$)} & 4.30 & 2.95 & 2.19 & 0.017 & 0.9 & 1.43 & 1.48 & 3.40 ($\pm 0.04$) & \posg{+8\%}\\
               & Cov@100 & \underline{19.73 ($\pm 0.68$)} & 0.06 & 0.08 & 14.58 & 96.05 & 16.70 & 46.11 & 47.17 & 60.03 ($\pm 1.31$) & \posg{+204\%}\\
    \bottomrule
  \end{tabular}
\end{table*}

To compare our approach with the modern multi-representational models in consistent way, we utilized the setup proposed in the work LOOM \cite{loom} using the SASRec \cite{scalablece} model as a bridge between two setups. 
As could be seen in the Table \ref{tab:loom}, our approach significantly outperforms other multi-representational models, such as LOOM \cite{loom}, Self-GNN \cite{liu2024selfgnn} and GSAU \cite{gsau}.


\section{Ablation study}
\label{ablation}
In this section, we ablate various components of our model and how they affect the quality of recommendations. The parameters of interest were the number of epochs for warming up the graph encoder, the size of the graph input to our model, various representation alignment methods, and different backbones for the sequential and graph encoders. The results are presented below.

\begin{table}[ht]
\caption{Model comparison in LOOM \cite{loom} setup using $HR@10$ and $NDCG@10$ metrics (in $\%$).}
\centering
\setlength{\tabcolsep}{5pt} 
\renewcommand{\arraystretch}{1.4} 
\begin{tabular}{l l c c c c c}
\toprule
\rotatebox{55}{Dataset} & \rotatebox{55}{Metric} & 
\rotatebox{55}{SASRec} & 
\rotatebox{55}{GSAU} & 
\rotatebox{55}{Self-GNN} & 
\rotatebox{55}{LOOM} & 
\rotatebox{55}{CREATE} \\
\midrule
\rotatebox{0}{Office} & HR@10 & 1.81 & 1.95 & 1.42 & \underline{2.11} & \textbf{2.16} \\
       & NDCG@10 & 0.72 & 0.78 & 0.55 & \underline{0.92} & \textbf{0.96} \\
\midrule
\rotatebox{0}{Beauty} & HR@10 & 1.19 & 1.35 & 0.98 & \underline{1.43} & \textbf{1.49} \\
       & NDCG@10 & 0.49 & 0.54 & 0.40 & \underline{0.55} & \textbf{0.57} \\
\bottomrule
\end{tabular}
\label{tab:loom}
\end{table}

\subsection{Graph size}
To have a global understanding of users' interactions, we first build user-item graph and then learn it. Evidently, the information contained in the graph influences the resulting performance of the framework. Thus, it is essential to explore the possibilities for further development. In this section, we examine the influence of graph size on overall model performance on different datasets.

\begin{figure}[h]
    \centering
    \includegraphics[width=0.48\textwidth]{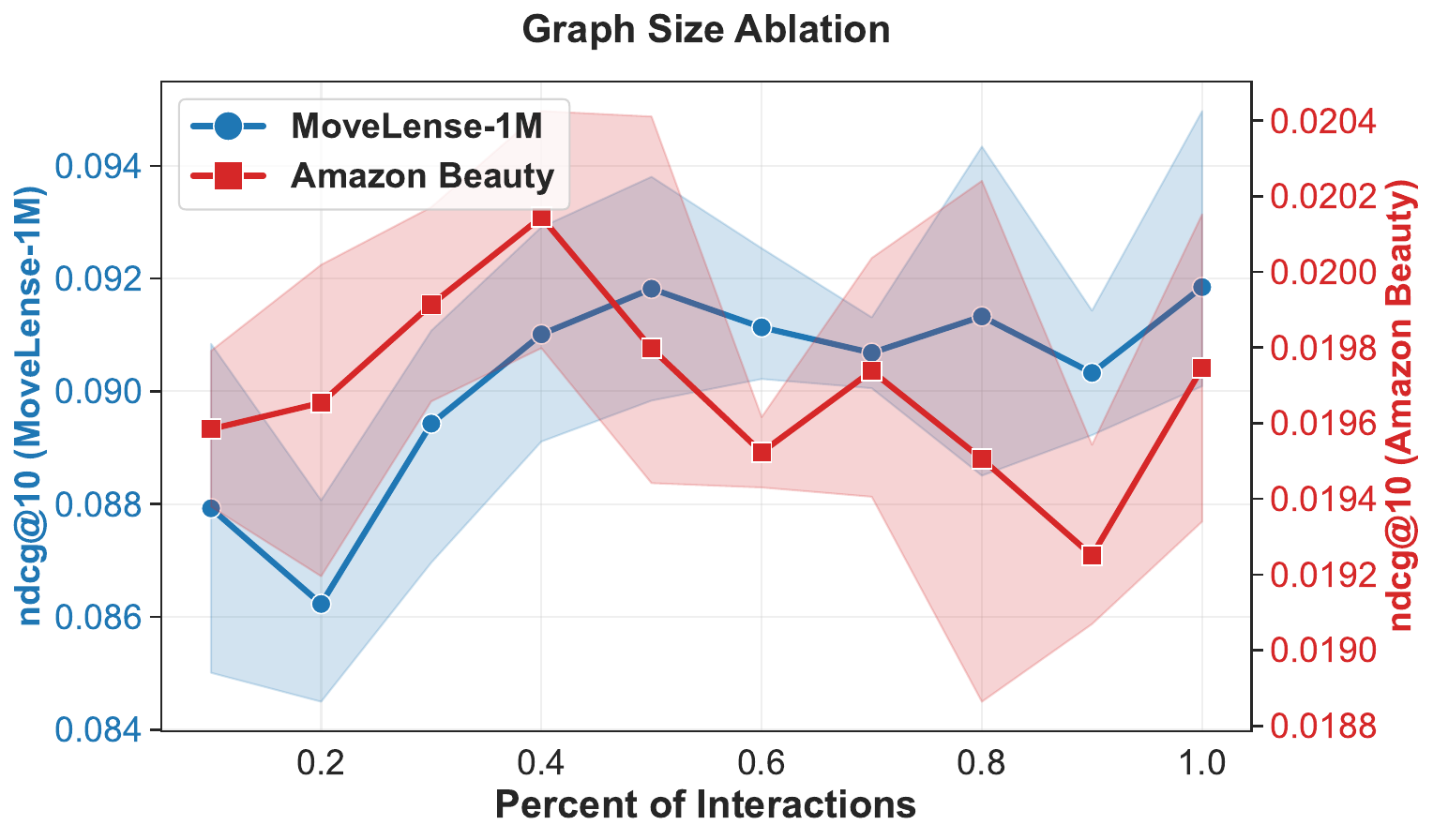}
    \caption{Graph size ablation for MovieLens-1M and Amazon Beauty datasets}
    \Description{The plot shows NDCG@10 score for each graph built on $n\%$ of latest interactions per user with $n\%$ ranging from $10\%$ to $100\%$. Primary axis depicts the scale for MovieLens-1M and secondary axis corresponds for the values of NDCG@10 on Amazon Beauty. Amazon Beauty has optimal values around $40\%$, while MovieLens-1M has optimal values $n > 40\%$.}
    \label{fig:graph_size_ablation}
\end{figure}

The default way to build user-item interactions graph is to take $N$ last user interactions, where $N$ is a hyperparameter for each dataset. In this section, we conduct the experiments by taking different percent $n\%$ of the last interactions per user ranging from $10\%$ to $100\%$. For each setup, we took the optimal hyperparameter configuration and measured the mean NDCG@10 on four different seeds. Each dataset was split by global temporal split into $80\%$ for the train set and $20\%$ for the test set. In a side-by-side comparison of experiments conducted on MovieLens-1M and Amazon Beauty (Figure \ref{fig:graph_size_ablation}), it is evident that each dataset has an optimal percent of interactions, leading to graphs with richer structure and better performance. For MovieLens-1M all graphs that cover more than $40\%$ of interactions lead to better performance, while for Amazon Beauty dataset the optimal value is approximately $40\%$. Lower NDCG@10 scores on graphs with greater amount of interactions may be caused by insufficient number of warm-up epochs, which help to process complex graph structures.

\subsection{Training with GCN warm-up}
The warm-up phase pre-trains the graph encoder so that its representations do not destabilize the sequential encoder during early training. Its required length is data-dependent: it can be skipped on simpler graphs and extended on more structured ones. We start from the plain sequential setting and sweep the number of warm-up epochs from \(0\) to \(100\); we report mean NDCG@10 over \(3\) seeds using a global temporal split.

\begin{figure}[h]
    \centering
    \includegraphics[width=0.47\textwidth]{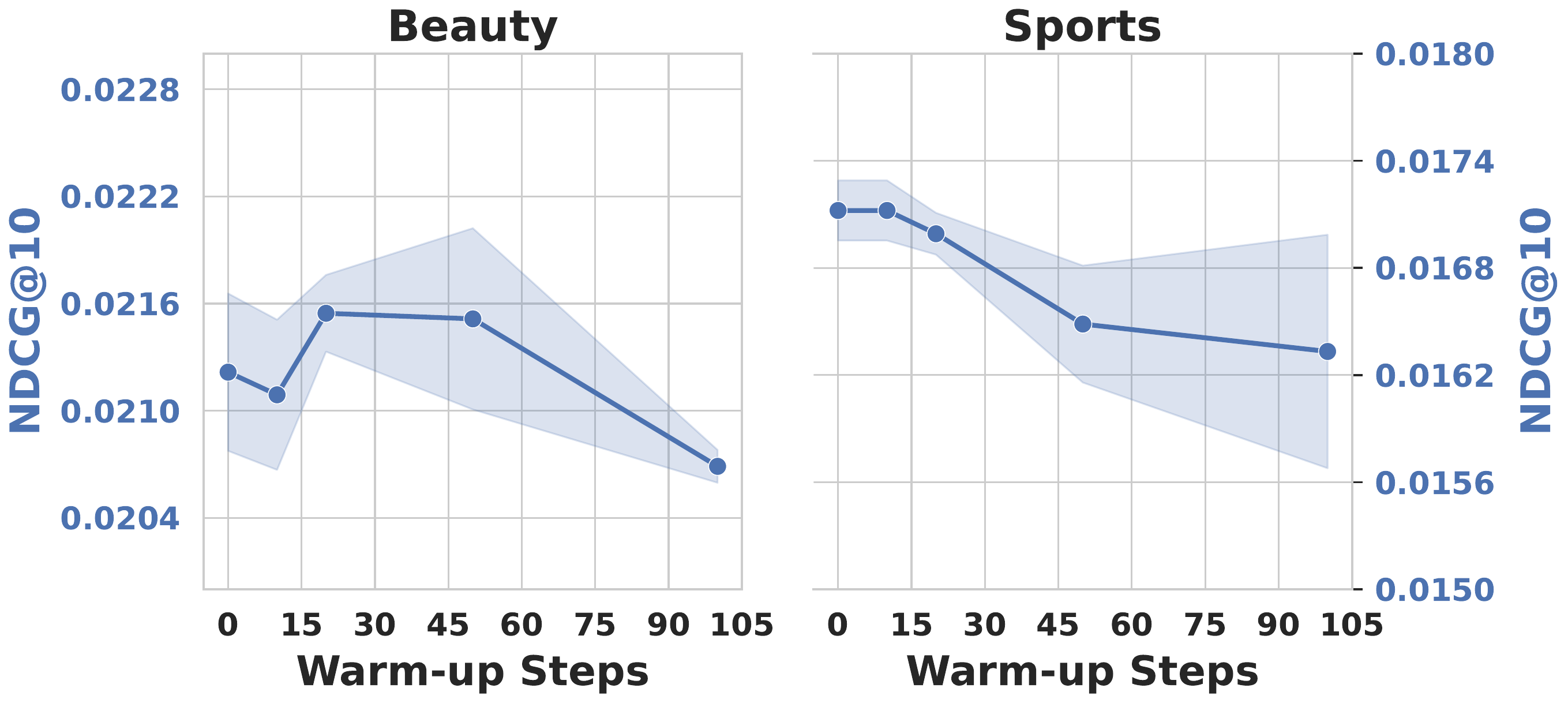}
    \caption{Performance (NDCG@10) vs.\ the number of GCN warm-up epochs.}
    \Description{Line plot of NDCG@10 on Amazon Beauty and Sports versus GCN warm-up epochs; performance peaks on  50 for Beauty and 10 epochs for Sports and declines with longer warm-up.}
    \label{fig:warm_up_ablation}
\end{figure}

We evaluate this procedure on Amazon Beauty and Sports. For each dataset, we measure the best test NDCG@10 over three seeds for warm-up lengths of \(0,10,20,50,\) and \(100\) epochs. In both datasets, warm-up improves performance over no warm-up, with the best results at the \(50\) and \(10\)  epochs correspondingly which confirms the choice of the hyperparameter given by Optuna (Fig.~\ref{fig:warm_up_ablation}). Longer warm-up can overfit the graph encoder and reduce end-to-end performance, so the number of warm-up epochs should be tuned.

\subsection{Representation alignment}
Despite the effectiveness of both sequential and graph recommender systems, they solve different tasks and their representations of data differ from each other. Moreover, while sequential models embed only items \cite{sasrec}, the graph model \cite{he2020lightgcn} represents both users and items in the vector space. Additionally, as shown in Table \ref{tab:ablation}, the performance of the multi-representational model without graph-sequential alignment is lower than that of the sequential model.

\begin{figure}
    \centering
    \includegraphics[width=0.47\textwidth]{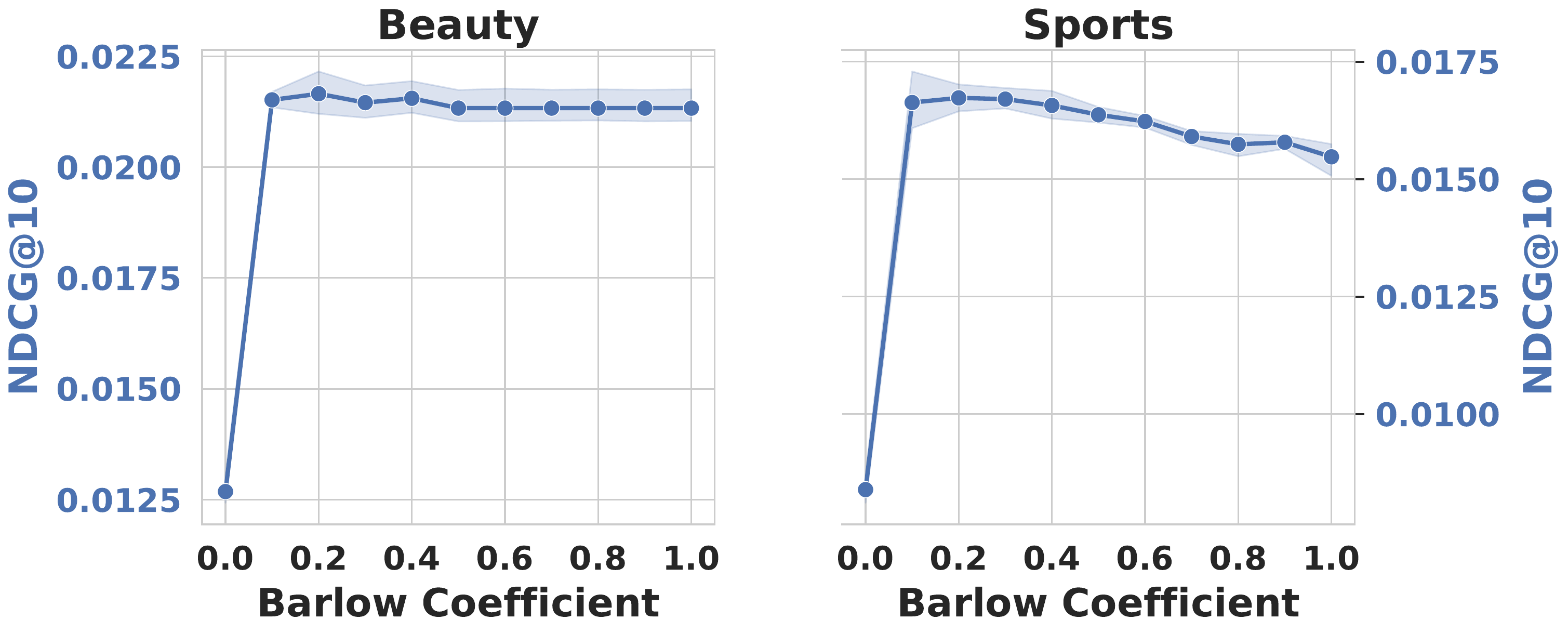}
    \caption{Test ndcg@10 metric by the Barlow coefficient.}
    \Description{Line plots showing the impact of the Barlow Twins loss coefficient on NDCG@10 for Beauty and Sports datasets. Both plots show a sharp increase in performance as the coefficient moves from 0.0 to 0.1, confirming that representation alignment significantly improves recommendation quality. For the Beauty dataset, performance remains stable around 0.0215 as the coefficient increases from 0.1 to 1.0. For the Sports dataset, performance peaks around 0.0165 at a coefficient of 0.2 and then slightly declines to approximately 0.0155 as the coefficient approaches 1.0. This demonstrates that while the Barlow Twins loss is crucial, its optimal weight varies slightly by dataset.}

    \label{fig:barlow}
\end{figure}

To address these issues, some works—such as \cite{mrgsrec,yusupov2026efficient}—employ contrastive learning, while others propose the Barlow Twins approach \cite{razvorotnev2025barlow}. We evaluate our method in three settings: without representation alignment, with the contrastive loss, and with the Barlow Twins loss. Our ablation study shows that the Barlow Twins approach significantly improves recommendation quality, albeit at the cost of reduced coverage metrics. Across alignment losses, UltraGCN achieves superior performance over LightGCN ones in $K=10$ quality metrics, whilst the improvements for $K=100$ are not significant. The benefit of the Barlow Twins approach \cite{razvorotnev2025barlow} can be explained by the fact that it does not rely on negative sampling techniques, which often do not fully capture the real data distribution. Moreover, the Barlow Twins approach tends to recommend rarer but relevant items to users, leading to improved recommendation metrics. The correspondence of the test NDCG@10 metric to Barlow Twins coefficient is plotted on Figure~\ref{fig:barlow}.

\begin{table}[ht]
\centering
\caption{Comparison of our model with different alignment losses and the sequential baseline on the Sports and Clothing datasets, using $NDCG@K$ and $Coverage@K$ metrics (in $\%$).}
\begin{tabular}{lcccc}
\toprule
\multicolumn{5}{c}{\textbf{Sports}} \\
\midrule
Method & NDCG@10 & Cov@10 & NDCG@100 & Cov@100 \\
\midrule
Sequential & 1.48 & 50.35 & 3.04 & 79.41 \\
\midrule
\multicolumn{5}{c}{\textit{CREATE with LightGCN graph encoder}} \\
\midrule
None   & 1.37 & 36.58 & 2.89 & 66.95 \\
Contrastive & 1.49 & \textbf{40.16} & 3.05 & \textbf{67.59} \\
Barlow & \textbf{1.56} & 25.69 & \textbf{3.24} & 66.02 \\
\midrule
\multicolumn{5}{c}{\textit{CREATE with UltraGCN graph encoder}} \\
\midrule
None   & 1.38 & 44.32 & 2.86 & 89.76 \\
Contrastive & 1.48 & 60.39 & 3.07 & 87.19 \\
Barlow & \textbf{1.58} & \textbf{75.90} & \textbf{3.24} & \textbf{96.33} \\
\midrule
\midrule
\multicolumn{5}{c}{\textbf{Clothing}} \\
\midrule
Method & N@10 & C@10 & N@100 & C@100 \\
\midrule
Sequential & 0.87 & 52.11 & 1.86 & 84.55 \\
\midrule
\multicolumn{5}{c}{\textit{CREATE with LightGCN graph encoder}} \\
\midrule
None   & 0.90 & \textbf{77.71} & 1.87 & \textbf{94.37} \\
Contrastive & 0.92 & 69.53 & 1.89 & 89.17 \\
Barlow & \textbf{0.95} & 72.48 & \textbf{1.92} & 93.02 \\
\midrule
\multicolumn{5}{c}{\textit{CREATE with UltraGCN graph encoder}} \\
\midrule
None   & 0.92 & 53.79 & 2.05 & 85.34 \\
Contrastive & 0.95 & 48.84 & 2.10 & 83.54 \\
Barlow & \textbf{1.00} & \textbf{57.06} & \textbf{2.19} & \textbf{87.67} \\
\bottomrule
\end{tabular}
\label{tab:ablation}
\end{table}


\subsection{Using different encoders}

Our CREATE approach effectively combines sequential models like SASRec \cite{scalablece} and BERT4Rec \cite{bert4rec} with graph-based models like LightGCN \cite{he2020lightgcn} and UltraGCN \cite{mao2021ultragcn}. It outperforms both pure sequential and pure graph baselines (see Tables \ref{table:comparison_full}, \ref{tab:ablation} and \ref{tab:encoders}). This improvement can be explained by the fact that integrating sequential and graph-based patterns provides a richer representation of user behavior than either pattern type alone. 

Moreover, the performance of CREATE improves with the improvement of the underlying sequential and graph encoder quality: the stronger the graph component, the better the overall results. Since our approach works with various sequential and graph encoders, it is highly flexible and can be improved by plugging in enhanced sequential or graph components.

An additional advantage of CREATE is that it does not require user embeddings during inference. Consequently, there is no need to develop folding-in methods for its graph components \cite{yusupov2025ultra,yusupov2026efficient}, which simplifies the model pipeline.


\begin{table}[ht]
\caption{Comparison of sequential backbones (SASRec, BERT4Rec) combined with graph encoders (LightGCN, UltraGCN) evaluated by $NDCG@10$ and $Recall@10$.}
\centering
\setlength{\tabcolsep}{5pt} 
\renewcommand{\arraystretch}{1.4} 
\begin{tabular}{l l  c c }
\toprule
\rotatebox{35}{Seq. Backbone} & \rotatebox{35}{Metric} & 
\rotatebox{35}{LightGCN} & 
\rotatebox{35}{UltraGCN} \\
\midrule
SASRec & NDCG@10  & 1.67 & 1.69 \\
       & Recall@10  & 3.02 & 3.13  \\
\midrule
BERT4Rec & NDCG@10  & 1.39 & 1.43  \\
       & Recall@10  & 2.65 & 2.73  \\
\bottomrule
\end{tabular}
\label{tab:encoders}
\end{table}

\section{Conclusion}

\label{conclusion}
This paper presents a framework for training recommender models that combines sequential and graph representations of user interactions. This enables the enrichment of vector representations of objects with both local and global information for next-item prediction. A method for training a model within this framework was developed, along with a method for representation alignment between encoders, which, improves the quality of recommendations. Experiments demonstrate improved performance of the proposed method compared to both classical baseline approaches and modern sequential models, as well as recent work in the same field under realistic conditions. An ablation study confirms the influence of various components and parameters provided by the framework on the final quality of recommendations.


\bibliographystyle{ACM-Reference-Format}
\bibliography{sample-base}

@String{Computing = "Computing" }

@String{Springer = "Springer-Verlag" }

@misc{sasrec,
      title={Self-Attentive Sequential Recommendation}, 
      author={Wang-Cheng Kang and Julian McAuley},
      year={2018},
      eprint={1808.09781},
      archivePrefix={arXiv},
      primaryClass={cs.IR},
      url={https://arxiv.org/abs/1808.09781}, 
}

@misc{bert4rec,
      title={BERT4Rec: Sequential Recommendation with Bidirectional Encoder Representations from Transformer}, 
      author={Fei Sun and Jun Liu and Jian Wu and Changhua Pei and Xiao Lin and Wenwu Ou and Peng Jiang},
      year={2019},
      eprint={1904.06690},
      archivePrefix={arXiv},
      primaryClass={cs.IR},
      url={https://arxiv.org/abs/1904.06690}, 
}

@inproceedings{scalablece, series={RecSys ’24},
   title={Scalable Cross-Entropy Loss for Sequential Recommendations with Large Item Catalogs},
   url={http://dx.doi.org/10.1145/3640457.3688140},
   DOI={10.1145/3640457.3688140},
   booktitle={18th ACM Conference on Recommender Systems},
   publisher={ACM},
   author={Mezentsev, Gleb and Gusak, Danil and Oseledets, Ivan and Frolov, Evgeny},
   year={2024},
   month=oct, pages={475–485},
   collection={RecSys ’24} }

@inproceedings{gusak2025time,
  title={Time to split: Exploring data splitting strategies for offline evaluation of sequential recommenders},
  author={Gusak, Danil and Volodkevich, Anna and Klenitskiy, Anton and Vasilev, Alexey and Frolov, Evgeny},
  booktitle={Proceedings of the Nineteenth ACM Conference on Recommender Systems},
  pages={874--883},
  year={2025}
}

@inproceedings{he2020lightgcn,
  title={Lightgcn: Simplifying and powering graph convolution network for recommendation},
  author={He, Xiangnan and Deng, Kuan and Wang, Xiang and Li, Yan and Zhang, Yongdong and Wang, Meng},
  booktitle={Proceedings of the 43rd International ACM SIGIR conference on research and development in Information Retrieval},
  pages={639--648},
  year={2020}
}

@inproceedings{mao2021ultragcn,
  title={UltraGCN: ultra simplification of graph convolutional networks for recommendation},
  author={Mao, Kelong and Zhu, Jieming and Xiao, Xi and Lu, Biao and Wang, Zhaowei and He, Xiuqiang},
  booktitle={Proceedings of the 30th ACM international conference on information \& knowledge management},
  pages={1253--1262},
  year={2021}
}

@inproceedings{loom,
  title={Linking Ordered and Orderless Modeling for Sequential Recommendation},
  author={Li, Sijia and Gao, Min and Wang, Zongwei and Bai, Yibing and Chen, Wuhan},
  booktitle={Proceedings of the 34th ACM International Conference on Information and Knowledge Management},
  pages={1654--1664},
  year={2025}
}

@inproceedings{gsau,
  title={Graph-Sequential Alignment and Uniformity: Toward Enhanced Recommendation Systems},
  author={Cao, Yuwei and Yang, Liangwei and Liu, Zhiwei and Liu, Yuqing and Wang, Chen and Liang, Yueqing and Peng, Hao and Yu, Philip S},
  booktitle={Companion Proceedings of the ACM on Web Conference 2025},
  pages={888--892},
  year={2025}
}

@article{razvorotnev2025barlow,
  title={Barlow Twins for Sequential Recommendation},
  author={Razvorotnev, Ivan and Munkhoeva, Marina and Frolov, Evgeny},
  journal={arXiv preprint arXiv:2510.26407},
  year={2025}
}

@inproceedings{BPRMF,
	author    = {Steffen Rendle and
	Christoph Freudenthaler and
	Zeno Gantner and
	Lars Schmidt{-}Thieme},
	title     = {{BPR:} Bayesian Personalized Ranking from Implicit Feedback},
	booktitle = {{UAI}},
	pages     = {452--461},
	year      = {2009}
}

@misc{asymmetrichyperbolic,
      title={Scalable Hyperbolic Recommender Systems}, 
      author={Benjamin Paul Chamberlain and Stephen R. Hardwick and David R. Wardrope and Fabon Dzogang and Fabio Daolio and Saúl Vargas},
      year={2019},
      eprint={1902.08648},
      archivePrefix={arXiv},
      primaryClass={cs.IR},
      url={https://arxiv.org/abs/1902.08648}, 
}

@inproceedings{yambda, author = {Ploshkin, Alexander and Tytskiy, Vladislav and Pismenny, Alexey and Baikalov, Vladimir and Taychinov, Evgeny and Permiakov, Artem and Burlakov, Daniil and Krofto, Eugene}, title = {Yambda-5B — A Large-Scale Multi-Modal Dataset for Ranking and Retrieval}, year = {2025}, isbn = {9798400713644}, publisher = {Association for Computing Machinery}, address = {New York, NY, USA}, url = {https://doi.org/10.1145/3705328.3748163}, doi = {10.1145/3705328.3748163}, booktitle = {Proceedings of the Nineteenth ACM Conference on Recommender Systems}, pages = {894–901}, numpages = {8}, location = { }, series = {RecSys '25} }

@misc{amazon_review_data_jmcauley,
    title = {Amazon Review Data},
    url = {https://jmcauley.ucsd.edu/data/amazon/},
    author = {McAuley, J.},
    year = {2016}
}

@inproceedings{zbontar2021barlowtwins,
  title     = {Barlow Twins: Self-Supervised Learning via Redundancy Reduction},
  author    = {Zbontar, Jure and Jing, Li and Misra, Ishan and LeCun, Yann and Deny, St{\'e}phane},
  booktitle = {Proceedings of the 38th International Conference on Machine Learning (ICML)},
  series    = {Proceedings of Machine Learning Research},
  volume    = {139},
  pages     = {12310--12320},
  year      = {2021},
  publisher = {PMLR},
  url       = {https://proceedings.mlr.press/v139/zbontar21a.html}
}

@article{yusupov2026efficient,
  title={Efficient incorporation of new interactions in graph recommenders via folding-in: V. Yusupov et al.},
  author={Yusupov, Viacheslav and Sukhorukov, Nikita and Frolov, Evgeny},
  journal={User Modeling and User-Adapted Interaction},
  volume={36},
  number={1},
  pages={2},
  year={2026},
  publisher={Springer}
}

@article{zhang2019deep,
  title={Deep learning based recommender system: A survey and new perspectives},
  author={Zhang, Shuai and Yao, Lina and Sun, Aixin and Tay, Yi},
  journal={ACM computing surveys (CSUR)},
  volume={52},
  number={1},
  pages={1--38},
  year={2019},
  publisher={ACM New York, NY, USA}
}

@article{gheewala2025depth,
  title={In-depth survey: deep learning in recommender systems—exploring prediction and ranking models, datasets, feature analysis, and emerging trends},
  author={Gheewala, Shivangi and Xu, Shuxiang and Yeom, Soonja},
  journal={Neural Computing and Applications},
  pages={1--73},
  year={2025},
  publisher={Springer}
}

@article{yoon2023evolution,
  title={Evolution of deep learning-based sequential recommender systems: From current trends to new perspectives},
  author={Yoon, Ji Hyung and Jang, Beakcheol},
  journal={IEEE Access},
  volume={11},
  pages={54265--54279},
  year={2023},
  publisher={IEEE}
}

@article{kipf2016semi,
  title={Semi-supervised classification with graph convolutional networks},
  author={Kipf, TN},
  journal={arXiv preprint arXiv:1609.02907},
  year={2016}
}

@article{wu2022graph,
  title={Graph neural networks in recommender systems: a survey},
  author={Wu, Shiwen and Sun, Fei and Zhang, Wentao and Xie, Xu and Cui, Bin},
  journal={ACM Computing Surveys},
  volume={55},
  number={5},
  pages={1--37},
  year={2022},
  publisher={ACM New York, NY}
}

@article{anand2025survey,
  title={A survey on recommender systems using graph neural network},
  author={Anand, Vineeta and Maurya, Ashish Kumar},
  journal={ACM Transactions on Information Systems},
  volume={43},
  number={1},
  pages={1--49},
  year={2025},
  publisher={ACM New York, NY, USA}
}

@inproceedings{li2020time,
  title={Time interval aware self-attention for sequential recommendation},
  author={Li, Jiacheng and Wang, Yujie and McAuley, Julian},
  booktitle={Proceedings of the 13th international conference on web search and data mining},
  pages={322--330},
  year={2020}
}

@article{lichtenberg2025denserec,
  title={DenseRec: Revisiting Dense Content Embeddings for Sequential Transformer-based Recommendation},
  author={Lichtenberg, Jan Malte and De Candia, Antonio and Ruffini, Matteo},
  journal={arXiv preprint arXiv:2508.18442},
  year={2025}
}

@inproceedings{xie2022contrastive,
  title={Contrastive learning for sequential recommendation},
  author={Xie, Xu and Sun, Fei and Liu, Zhaoyang and Wu, Shiwen and Gao, Jinyang and Zhang, Jiandong and Ding, Bolin and Cui, Bin},
  booktitle={2022 IEEE 38th international conference on data engineering (ICDE)},
  pages={1259--1273},
  year={2022},
  organization={IEEE}
}

@inproceedings{wang2019neural,
  title={Neural graph collaborative filtering},
  author={Wang, Xiang and He, Xiangnan and Wang, Meng and Feng, Fuli and Chua, Tat-Seng},
  booktitle={Proceedings of the 42nd international ACM SIGIR conference on Research and development in Information Retrieval},
  pages={165--174},
  year={2019}
}

@article{kim2025graph,
  title={Graph-based technology recommendation system using GAT-NGCF},
  author={Kim, Min-Seung and Jang, Yong-Ju and Sung, Tae-Eung},
  journal={Expert Systems with Applications},
  pages={128240},
  year={2025},
  publisher={Elsevier}
}

@inproceedings{mrgsrec,
  title={End-to-end graph-sequential representation learning for accurate recommendations},
  author={Baikalov, Vladimir and Frolov, Evgeny},
  booktitle={Companion Proceedings of the ACM Web Conference 2024},
  pages={501--504},
  year={2024}
}

@inproceedings{wang2022multi,
  title={Multi-level contrastive learning framework for sequential recommendation},
  author={Wang, Ziyang and Liu, Huoyu and Wei, Wei and Hu, Yue and Mao, Xian-Ling and He, Shaojian and Fang, Rui and Chen, Dangyang},
  booktitle={Proceedings of the 31st ACM international conference on information \& knowledge management},
  pages={2098--2107},
  year={2022}
}

@inproceedings{yusupov2025ultra,
  title={Ultra Fast Warm Start Solution for Graph Recommendations},
  author={Yusupov, Viacheslav and Rakhuba, Maxim and Frolov, Evgeny},
  booktitle={Proceedings of the 34th ACM International Conference on Information and Knowledge Management},
  pages={5469--5473},
  year={2025}
}

@inproceedings{klenitskiy2024does,
  title={Does it look sequential? an analysis of datasets for evaluation of sequential recommendations},
  author={Klenitskiy, Anton and Volodkevich, Anna and Pembek, Anton and Vasilev, Alexey},
  booktitle={Proceedings of the 18th ACM Conference on Recommender Systems},
  pages={1067--1072},
  year={2024}
}

@inproceedings{du2024disentangled,
  title={Disentangled Multi-interest Representation Learning for Sequential Recommendation},
  author={Du, Yingpeng and Wang, Ziyan and Sun, Zhu and Ma, Yining and Liu, Hongzhi and Zhang, Jie},
  booktitle={Proceedings of the 30th ACM SIGKDD Conference on Knowledge Discovery and Data Mining},
  pages={677--688},
  year={2024}
}

@inproceedings{zhang2024temporal,
  title={Temporal graph contrastive learning for sequential recommendation},
  author={Zhang, Shengzhe and Chen, Liyi and Wang, Chao and Li, Shuangli and Xiong, Hui},
  booktitle={Proceedings of the AAAI conference on artificial intelligence},
  volume={38},
  number={8},
  pages={9359--9367},
  year={2024}
}

@article{zare2025dac,
  title={DAC-GCN: A Dual Actor-Critic Graph Convolutional Network with Multi-Hop Aggregation for Enhanced Recommender Systems},
  author={Zare, Gholamreza and Jafari, Nima and Hosseinzadeh, Mehdi and Sahafi, Amir},
  journal={Acta Informatica Pragensia},
  volume={2025},
  number={3},
  pages={340--364},
  year={2025},
  publisher={Prague University of Economics and Business}
}

@article{yu2023xsimgcl,
  title={XSimGCL: Towards extremely simple graph contrastive learning for recommendation},
  author={Yu, Junliang and Xia, Xin and Chen, Tong and Cui, Lizhen and Hung, Nguyen Quoc Viet and Yin, Hongzhi},
  journal={IEEE Transactions on Knowledge and Data Engineering},
  volume={36},
  number={2},
  pages={913--926},
  year={2023},
  publisher={IEEE}
}

@inproceedings{chen2025lightgnn,
  title={Lightgnn: Simple graph neural network for recommendation},
  author={Chen, Guoxuan and Xia, Lianghao and Huang, Chao},
  booktitle={Proceedings of the Eighteenth ACM International Conference on Web Search and Data Mining},
  pages={549--558},
  year={2025}
}

@inproceedings{zhang2024transgnn,
  title={Transgnn: Harnessing the collaborative power of transformers and graph neural networks for recommender systems},
  author={Zhang, Peiyan and Yan, Yuchen and Zhang, Xi and Li, Chaozhuo and Wang, Senzhang and Huang, Feiran and Kim, Sunghun},
  booktitle={Proceedings of the 47th International ACM SIGIR conference on research and development in information retrieval},
  pages={1285--1295},
  year={2024}
}

@article{luo2024collaborative,
  title={Collaborative sequential recommendations via multi-view gnn-transformers},
  author={Luo, Tianze and Liu, Yong and Pan, Sinno Jialin},
  journal={ACM Transactions on Information Systems},
  volume={42},
  number={6},
  pages={1--27},
  year={2024},
  publisher={ACM New York, NY}
}

@inproceedings{ho2024self,
  title={Self-Attentive Sequential Recommendation Models Enriched with More Features},
  author={Ho, Trong Dang Huu and Nguyen, Sang Thi Thanh},
  booktitle={Proceedings of the 2024 8th International Conference on Deep Learning Technologies (ICDLT)},
  pages={49--55},
  year={2024}
}

@inproceedings{rashed2022context,
  title={Context and attribute-aware sequential recommendation via cross-attention},
  author={Rashed, Ahmed and Elsayed, Shereen and Schmidt-Thieme, Lars},
  booktitle={Proceedings of the 16th ACM conference on recommender systems},
  pages={71--80},
  year={2022}
}

@inproceedings{yuan2025contextual,
  title={A Contextual-Aware Position Encoding for Sequential Recommendation},
  author={Yuan, Jun and Cai, Guohao and Dong, Zhenhua},
  booktitle={Companion Proceedings of the ACM on Web Conference 2025},
  pages={577--585},
  year={2025}
}

@article{boka2024survey,
  title={A survey of sequential recommendation systems: Techniques, evaluation, and future directions},
  author={Boka, Tesfaye Fenta and Niu, Zhendong and Neupane, Rama Bastola},
  journal={Information Systems},
  volume={125},
  pages={102427},
  year={2024},
  publisher={Elsevier}
}

@article{gao2023survey,
  title={A survey of graph neural networks for recommender systems: Challenges, methods, and directions},
  author={Gao, Chen and Zheng, Yu and Li, Nian and Li, Yinfeng and Qin, Yingrong and Piao, Jinghua and Quan, Yuhan and Chang, Jianxin and Jin, Depeng and He, Xiangnan and others},
  journal={ACM Transactions on Recommender Systems},
  volume={1},
  number={1},
  pages={1--51},
  year={2023},
  publisher={ACM New York, NY, USA}
}

@inproceedings{yang2023dgrec,
  title={Dgrec: Graph neural network for recommendation with diversified embedding generation},
  author={Yang, Liangwei and Wang, Shengjie and Tao, Yunzhe and Sun, Jiankai and Liu, Xiaolong and Yu, Philip S and Wang, Taiqing},
  booktitle={Proceedings of the sixteenth ACM international conference on web search and data mining},
  pages={661--669},
  year={2023}
}

@inproceedings{klenitskiy2023turning,
  title={Turning dross into gold loss: is bert4rec really better than sasrec?},
  author={Klenitskiy, Anton and Vasilev, Alexey},
  booktitle={Proceedings of the 17th ACM Conference on Recommender Systems},
  pages={1120--1125},
  year={2023}
}

@inproceedings{gusak2024rece,
  title={RECE: Reduced Cross-Entropy Loss for Large-Catalogue Sequential Recommenders},
  author={Gusak, Danil and Mezentsev, Gleb and Oseledets, Ivan and Frolov, Evgeny},
  booktitle={Proceedings of the 33rd ACM International Conference on Information and Knowledge Management},
  pages={3772--3776},
  year={2024}
}

@article{jing2023contrastive,
  title={Contrastive self-supervised learning in recommender systems: A survey},
  author={Jing, Mengyuan and Zhu, Yanmin and Zang, Tianzi and Wang, Ke},
  journal={ACM Transactions on Information Systems},
  volume={42},
  number={2},
  pages={1--39},
  year={2023},
  publisher={ACM New York, NY}
}

@inproceedings{liu2024selfgnn,
  title={Selfgnn: Self-supervised graph neural networks for sequential recommendation},
  author={Liu, Yuxi and Xia, Lianghao and Huang, Chao},
  booktitle={Proceedings of the 47th international ACM SIGIR conference on research and development in information retrieval},
  pages={1609--1618},
  year={2024}
}


\end{document}